\documentclass[]{tPHM2e}

\usepackage{amssymb}
\usepackage{ctable}
\usepackage{tabularx}
\usepackage{multirow}

\begin{document}
\doi{10.1080/14786435.2010.521530}
\issn{1478-6443}
\issnp{1478-6435}
\jvol{00} \jnum{00} \jyear{2008} \jmonth{21 December}

\markboth{Ottochian \& Leporini}{Philosophical Magazine}

\articletype{ARTICLE}

\title{Scaling between structural relaxation and caged
  dynamics \\ in Ca${}_{\, 0.4}$K${}_{\, 0.6}$(NO${}_{\,
    3}$)${}_{\,1.4}$ and glycerol: free volume, time scales and
  implications for the pressure-energy correlations} 

\author{A.Ottochian and
  D. Leporini$^{\ast}$\thanks{$^\ast$Corresponding author. Email:
    dino.leporini@df.unipi.it
    \vspace{6pt}}\\\vspace{6pt} {\em{Dipartimento di Fisica ``Enrico
      Fermi'', \\ Universit\`{a} di Pisa, Largo B. Pontecorvo 3,
      I-56127 Pisa, Italy}}\\\vspace{6pt}\received{v4.4 released
    November 2008} }

\maketitle

\begin{abstract}
The scaling  of  the slow structural relaxation with the fast caged
dynamics is evidenced in the molten salt Ca${}_{\, 0.4}$K${}_{\,
  0.6}$(NO${}_{\, 3}$)${}_{\,1.4}$ (CKN) over about thirteen decades
of the structural relaxation time. Glycerol scaling was analyzed in
detail.  
In glycerol, the short-time mean-square displacement $\langle u^2
\rangle$, a measure of the caged dynamics, is contributed by
free-volume. It is seen that, in order to evidence the scaling, the
observation time of the fast dynamics must be shorter than the time
scales of the relaxation processes.
Systems with both negligible (like CKN, glycerol and network
glassformers) and high (like van der Waals liquids and polymers)
pressure-energy correlations exhibit the scaling between the slow
relaxation and the fast caged dynamics. According to the available
experiments, an isomorph-invariant expression of the master curve of
the scaled data is not distinguishable from a simpler not-invariant
expression. Instead, the latter agrees better with the simulations on
a wide class of model polymers.
\end{abstract}


\section{Introduction}
\label{sec:Introduction}
Understanding the extraordinary viscous slow-down that accompanies
glass formation is a major scientific challenge
\cite{Angell91,Angell95,DebeStilli2001}. On approaching the glass
transition (GT), trapping effects are more and more prominent.   The
average escape time from the cage of the first neighbors, i.e. the
structural relaxation time $\tau_\alpha$, increases from a few
picoseconds up to thousands of seconds. The rattling motion inside the
cage occurs on picosecond time scales with amplitude $\langle
u^2\rangle^{1/2}$. This quantity is related to the Debye-Waller factor
which, assuming harmonicity of thermal motion, takes the form
$\exp\left( -q^2 \langle u^2\rangle / 3 \right)$ where $q$ is the
absolute value of the scattering vector. At first sight, due to the
extreme time-scale separation between the rattling motion ($ \sim
10^{-12}\;\textrm{s}$) and the relaxation ($ \tau_\alpha \sim
10^2\textrm{s}$ at GT), one expects the complete independence of the
two motions. Nonetheless, several authors investigated their
correlations, emphasizing in particular the link with the bulk elastic
properties (for a review see ref. \cite{Dyre06}). In this research
field, the universal scaling between the structural relaxation time,
or the shear viscosity, and $\langle u^2\rangle$ was reported for
several numerical models, including linear polymers, mixtures,
prototypical glassformers like $SiO_2$ and o-terphenyl (OTP), and
icosahedral glassformer
\cite{OurNatPhys,OttochianEtAl09,OttochianLepoJNCS10}. The resulting
master curve fits with the available experimental data from
supercooled liquids, polymers and metallic glasses over about eighteen
decades of relaxation times (or viscosity in non-polymeric systems)
and a very wide range of fragilities.

The present paper shows that the scaling between the structural
relaxation time and $\langle u^2\rangle$ also holds for the molten
salt $\textrm{Ca}_{\, 0.4}$$\textrm{K}_{\,
  0.6}$$(\textrm{NO}_3)_{\,1.4}$ (CKN) over about thirteen decades of
$\tau_\alpha$. Glycerol scaling was analyzed in detail. The scaling of
glycerol and, over a less wide range of relaxation times and
viscosities, network glassformers like SiO${}_2$, GeO${}_2$ and
B${}_2$O${}_3$ was already reported
\cite{OurNatPhys,OttochianEtAl09,OttochianLepoJNCS10}. Here, for the
first time, the scaling of these systems and CKN will be discussed
from the viewpoint of the pressure-energy correlations
\cite{DyreJCP08,DyreJCP09,DyreIsomorphsJCP09}.
In addition, it is shown that the role of the free volume in glycerol
is not negligible and that the scaling works only if the observation
time of the fast dynamics is {\it shorter} than the time scales of the
relaxation processes \cite{OttochianLepoJNCS10}.

The paper first summarizes the results about the scaling between the
structural relaxation and $\langle u^2\rangle$. Then, the results on
glycerol and CKN are discussed in detail.  Finally, the conclusions
are presented. 

\section{Scaling between viscous flow, structural relaxation and
  $\langle u^2\rangle$} 
\label{sec:ModelAndTheory}
Fig.\ref{figure1} shows the scaling between the structural relaxation
time $\tau_\alpha$ and the viscosity $\eta$ with respect to the
reduced mean-square displacement (MSD), $\langle u^2 \rangle/ \langle
u^2_g \rangle$, with $ \langle u^2_g \rangle = \langle u^2
(T_g)\rangle$. The experimental data include supercooled liquids,
polymers, metallic glasses over about eighteen decades of relaxation
times and a very wide range of fragilities. The plot now includes the
ionic liquid CKN as well. All the data in Fig.\ref{figure1} are at
ambient pressure. The master curve of Fig.\ref{figure1} (black line)
is expressed analytically by:
\begin{equation}
  \label{eq:scaledparabola}
  \log X = \alpha + \tilde{\beta} \; \frac{\langle u^2_g
    \rangle}{\langle u^2 \rangle} + \tilde{\gamma} \left
  (\frac{\langle u^2_g \rangle}{\langle u^2\rangle} \right )^2
\end{equation}
with $X$ equal to the reduced quantities $\tau_\alpha/ \tau_0$ or
$\eta/ \eta_0 $. The best-fit values ( $\alpha=-0.424(1)$,
$\tilde{\beta}=1.62(6)$ and $\tilde{\gamma}=12.3(1)$ ) were drawn by
Molecular-Dynamics simulations (MD) on model polymeric systems
\cite{OurNatPhys} and mixtures, being confirmed by comparison with
prototypical glassformers like $\textrm {SiO}_2$ and o-terphenyl, and
icosahedral glassformer \cite{OttochianEtAl09}. For the systems in
Fig.\ref{figure1} Table \ref{tab:detailed} lists the MSD at the glass
transition temperature, i.e. the temperature where $\tau_{\alpha
}=10^2 \; \textrm{s}$ or $\eta =10^{12} \; \textrm{Pa $\cdot$ s} $,
and the conversion factors $\tau_0$ and $ \eta_0$ between the actual
time and viscosity units and the corresponding MD units,
respectively. Note that the conversion factors are the {\sl only}
adjustable parameters of the overall scaling procedure.
Table \ref{tab:detailed} shows also the (approximate) observation time
of the fast dynamics by the experiments $\Delta t$. The structural
relaxation times of the systems in Fig. \ref{figure1} are comparable
or longer than $\Delta t$. The $\Delta t$ dependence of the scaling of
glycerol is discussed in Sec. \ref{timescale}.

\section{Scaling and the pressure-energy correlations} 
\label{sec:u2fromPALS} 
Fig.\ref{figure2} shows that,  when the CKN structural relaxation time
and the glycerol shear viscosity are plotted vs the reduced MSD, they
collapse over about thirteen decades in a single master curve well
described by  Eq.\ref{eq:scaledparabola} within the error bars.  

There is competition between van der Waals and Coulombic terms in the
interacting  potential of CKN and glycerol. This feature deserves
consideration from the viewpoint of the pressure-energy correlations
\cite{DyreJCP08,DyreJCP09,DyreIsomorphsJCP09}. Such correlations are
expected to be weak for CKN and glycerol, thus leading to the absence
of isomorphic states in these two systems \cite{NoteAndalo}. Weak
correlations are also expected for network glassformers like
SiO${}_2$, GeO${}_2$ and B${}_2$O${}_3$  which exhibit scaling  (see
Fig.\ref{figure1}, not included in Fig.\ref{figure2} for clarity
reasons) \cite{OurNatPhys,OttochianEtAl09,OttochianLepoJNCS10}.  

Strongly correlating systems, e.g. van der Waals liquids,  have
isomorphic states. Rigorously,  any general theory of the liquid state
must end up in relations expressed only in terms of isomorph
invariants (constant quantities when evaluated over a set of
isomorphic states) to deal with the strongly correlating systems (a
criterion also known as the  ''isomorph filter''
\cite{DyreIsomorphsJCP09}). From this respect, the master curve given
by Eq.\ref{eq:scaledparabola}  with {\it constant} $\alpha,
\tilde{\beta}, \tilde{\gamma}$ parameters does {\it not} pass the
''isomorph filter'' and then should be unable to encompass strongly
correlating systems since: 
\begin{equation}
  \label{ratio_ISOMORPH_INV}
  \frac{\langle {\tilde{u}}^2_g
    \rangle}{\langle {\tilde{u}}^2 \rangle} \neq \frac{\langle u^2_g
    \rangle}{\langle u^2 \rangle}\hspace{1cm}, \hspace{1cm} \log \tilde{X} \neq  \log X
\end{equation}
where $\tilde{{\zeta}}  \equiv \zeta/\zeta_0$ with $\zeta_0 $ equal to
$\rho^{-1/3}, \rho^{-1/3} \sqrt{m/k_B T}, \rho^{2/3} \sqrt{m k_B T}$
for length, time and viscosity, respectively and $\tilde{X} =
\tilde{\tau_\alpha}/ \tilde{\tau_0}, \; \tilde{\eta}/ \tilde{\eta_0}$
\cite{NoteAndalo}. 
A proper isomorph-invariant modification of Eq.\ref{eq:scaledparabola}
is: 
\begin{equation}
  \log {\tilde{X}} = C_1 + C_2 \frac{\langle {\tilde{u}}_g^2
    \rangle}{\langle {\tilde{u}}^2 \rangle} + C_3  \left (
  \frac{\langle {\tilde{u}}_g^2 \rangle}{\langle {\tilde{u}}^2
    \rangle} \right )^2 
  \label{dyredensityquad}
\end{equation}
where $C_i, \; i=1,2,3$ are constants. The above equation is more
complex in nature than Eq. \ref{eq:scaledparabola} since it needs the
density as additional input parameter.   

The fact that Eq. \ref{dyredensityquad} is constrained by the basic
properties of strongly correlating systems, which are not accounted
for by Eq. \ref{eq:scaledparabola}, poses the question of the
flexibility of these master curves to deal with systems with either
strong (like OTP, TNB or polymers) or weak (like CKN, glycerol and the
network glassformers SiO${}_2$, GeO${}_2$ and B${}_2$O${}_3$)
pressure-energy correlations. May Eqs. \ref{eq:scaledparabola} and
\ref{dyredensityquad} be discriminated? To date, little help is
provided by the experiments. The experimental data plotted in
Fig.\ref{figure1} are taken at ambient pressure by sweeping the
temperature in a limited range, thus resulting in small density
changes. Owing to the much larger changes of the structural relaxation
time, the viscosity and MSD, the equalities $\log \tilde{X} \simeq
\log X$, $\langle {\tilde{u}}^2_g \rangle/ \langle {\tilde{u}}^2
\rangle \simeq \langle u^2_g  \rangle / \langle u^2 \rangle$ hold to a
good approximation and Eqs. \ref{eq:scaledparabola},
\ref{dyredensityquad} are hardly distinguishable. To our knowledge,
joint data concerning $\tau_\alpha$ (or viscosity for non polymeric
systems) and MSD from high-pressure experiments spanning larger
density changes are not available yet. If no clear-cut conclusion on
the experimental side may be still drawn, numerical studies offer more
insight. In particular,  Eqs. \ref{eq:scaledparabola} and
\ref{dyredensityquad} were investigated in the temperature/density
phase space of a class of  model polymers with different chain lengths
and generalized Lennard-Jones potentials having  constant position and
depth of their minimum \cite{OttochianEtAl09}.  
Note that pressure-energy correlations change with the potential and
the physical state \cite{CosloRolaJCP09}. 
The different systems were compared with no MSD or $\tau_\alpha$
rescaling and Eq.\ref{eq:scaledparabola} was used as master curve by
taking $\langle u^2_g  \rangle$ as a constant independent of the model
polymer. These positions: i) rely on the finding that the raw curves
$\tau_\alpha$ vs $\langle u^2 \rangle$ for different systems
superimpose very well, ii) fit in with the need to limit the number of
adjustable parameters. To keep on an equal footing the comparison
between the two master curves, an analogous position was adopted for
Eq. \ref{dyredensityquad}, i.e. $\langle {\tilde{u}}^2_g \rangle$ was
set to a constant independent of the model polymer (a step still
preserving the isomorph invariance of Eq.\ref{dyredensityquad}).  
On this basis,  Eq. \ref{eq:scaledparabola} was found to be more
consistent with the numerical data (see  Fig.11 of ref.
\cite{OttochianEtAl09}). To summarize, if the experiments do not
discriminate between the two versions of the master curve,  the
isomorph-invariant version Eq. \ref{dyredensityquad}, which is more
complex in nature than Eq. \ref{eq:scaledparabola} as noted above,
agrees less with  the simulations of a wide class of polymeric
systems.

\section{Free volume effects in Glycerol} 
\label{FreevolumeGLY} 
The closeness of $\langle u^2\rangle$ with free-volume concepts was
noted in experiments
\cite{KanayaEtAl99,SolesPRL01,DuvalEtAl97,DuvalEtAl98}, theories
\cite{HallWoly87,Ngai04,Dyre06} and simulations
\cite{OurNatPhys,OttochianEtAl09,Starr02} with some debate
\cite{Harrowell_NP08, AshtonGarraEPJE09}. { The basic question is
  about the existence of some critical displacement above which
  structural changes take place} and it was stated that the
rearrangement only takes place when there is a local, temporary
density decrease \cite{Dyre06}. We refer also to previous experimental
\cite{prevostoEtAlAl04,BarbieriEtAl04}, theoretical and simulation
\cite{BarbieriEtAl04bis, BarbieriEtAl04ter,AlessiJCP01,CristianoPRE01}
work.

The ratio between the volume that is accessible to the monomer
center-of-mass and the monomer volume is $v_0^{(MD)} \sim (2 \langle
u^2_g\rangle^{1/2})^3$ in MD units
\cite{OurNatPhys,OttochianEtAl09}. From our simulation data (first
line of Table \ref{tab:detailed}) one finds:
\begin{equation}
  v_0^{(MD)} \sim 0.017
  \label{freevol}
\end{equation}
\noindent
Flory and coworkers proposed that the glass transition takes place
under iso-free volume conditions with the universal value $v_0 \sim
0.025$ \cite{Gedde95}. This supports the conclusion that $\langle
u^2\rangle$  and the free volume are related, as above noted. 

To better elucidate the matter, we consider results from the Positron
Annihilation Lifetime Spectroscopy (PALS)
\cite{KanayaEtAl99,SolesPRL01,Bartos_Gly_JNCS05}. PALS is a structural
technique to parametrize the free volume. In fact, the
orthopositronium bound state of a positron has a strong tendency to
localize in holes of low electron density and decays with lifetime
$\tau_3$ being related to the average cavity size. In particular, we
test the ansatz :
\begin{equation}
  \langle u^2\rangle = C  \; \tau_3
  \label{u2proptotau3}
\end{equation}
where $C$ is a constant.  The above equation captures the essentials
of the relation between $\tau_3$ and $\langle u^2\rangle$, both
increasing with the free volume. Furthermore, it provides a simple
procedure to compare PALS and $\langle u^2\rangle$ data with no
adjustable parameters. In fact, Eq.\ref{u2proptotau3} yields the
equality $\langle u^2_g \rangle / \langle u^2 \rangle = \tau_{3
  g}/\tau_{3 } $, with $\tau_{3 g} = \tau_3 (T_g)$, i.e. the reduced
MSD is equal to the reduced PALS lifetime $\tau_3$. In
Fig. \ref{figure3} the viscosity data of glycerol were plotted by
replacing the reduced MSD with the reduced $\tau_3$ and comparing the
results to the master curve expressed by Eq. \ref{eq:scaledparabola}.
The conversion factor between the actual and the MD viscosity units
was adjusted to $\log \eta_0' = \log \eta_0 - 1$. Other parameters
like in Table \ref{tab:detailed}.  Fig. \ref{figure3} shows that the
scaling of the long-time dynamics by either the reduced $\tau_{3}$ or
the reduced MSD are quite close to each other over about ten decades
in relaxation times, thus leading to the conclusion that $\langle
u^2\rangle$ and free-volume are correlated. Deviations are observed
for glycerol when cage restructuring, and then the free-volume
fluctuations, occurs on time scales shorter than the PALS observation
time, i.e.  $\tau_\alpha \lesssim \tau_3 \sim 2\;\textrm{ns} $ (see
also ref.\cite{Bartos_Gly_JNCS05}). In this regime the cage dynamics
still takes place (it is detectable for $\tau_\alpha \gtrsim \; 1-10
\; \textrm{ps}$) and $\langle u^2\rangle$ is still well-defined
\cite{OurNatPhys,OttochianEtAl09} but the observation time $\tau_3$ is
too long and almost temperature-independent \cite{SolesPRL01}.

It must be noted that PALS data are usually interpreted by a model
assuming the spherical shape of the hole where the positron
decays. The model relates $\tau_{3}$ to the hole radius $R$
\cite{SolesPRL01}.  In addition to Eq.\ref{u2proptotau3}, we also
tested the relation $\langle u^2 \rangle \propto R^\delta$ ($R$
  drawn by the fit of $\tau_{3}$ data)  and found that the correlation
plot between $\log \eta$ and $R^\delta$ agrees with the master curve
expressed by Eq. \ref{eq:scaledparabola} with $\delta \sim
1$. However, being a model-dependent conclusion involving adjustable
parameters, we think that the analysis of the free-volume role in
terms of Eq. \ref{u2proptotau3} is more robust.

\section{Time scales of $\langle u^2 \rangle$ in Glycerol} 
\label{timescale} 

We now proceed and better clarify that the correct investigation of
the cage rattling is ensured if the observation time of the fast
dynamics is {\it shorter} than $\tau_\alpha$. Fig. \ref{figure4} shows
the correlation between $\langle u^2 \rangle$ and the viscosity data
of glycerol. $\langle u^2 \rangle$ was taken as the mean square
displacement within the time $\Delta t$, the latter depending on the
energy resolution of the neutron scattering (NS) experiment. It is
seen that the scaling between the long-time dynamics and $\langle u^2
\rangle_{\Delta t = 0.4 \; ns}$ is rather good even for $\tau_\alpha/
\Delta t \sim 1$, whereas using $\langle u^2 \rangle_{\Delta t = 5 \;
  ns}$ leads to deviations from the universal master curve if
$\tau_\alpha / \Delta t \lesssim 60$. This suggests that, if
$\tau_\alpha \lesssim 300 \; ns$, $\langle u^2 \rangle_{\Delta t = 5
  \; ns}$ is contributed by components which are not related to the
cage vibrational dynamics. One anticipates that these components have
relaxational character. Interestingly, $ \tau_3$ from PALS tracks the
universal master curve of the structural relaxation down to
$\tau_\alpha \sim 3 \; ns$ where $\tau_\alpha / \Delta t_{PALS} \sim
\tau_\alpha / \tau_3 \sim 1$ (Fig.\ref{figure3}). The finding seems to
indicate that in the range $3 \; ns \lesssim \tau_\alpha \lesssim 300
\; ns$ the free-volume is less affected by the relaxation than MSD on
the nanosecond time scale. The above discussion supports the
conclusion that cage-rattling MSD must be measured on time scales
$\Delta t$ shorter than the time scales due to both the structural and
the possible local relaxations. Shortening $\Delta t$ results in a
remarkable extension of the region where the scaling between the
structural relaxation and the cage dynamics of glycerol is observed
(Fig. \ref{figure4}). However, this effect is less apparent in OTP
\cite{OttochianLepoJNCS10}.

\section{Conclusions}
\label{sec:Conclusion}
It is shown that one molten salt (CKN) and one hydrogen-bonded liquid
(glycerol) exhibit the scaling between the mean-square displacement
due to cage rattling, $\langle u^2 \rangle$, and the structural
relaxation time or viscosity already observed in supercooled liquids,
polymers and metallic glasses. Systems with both negligible (like CKN,
glycerol, SiO${}_2$, GeO${}_2$ and B${}_2$O${}_3$) and high
pressure-energy correlations (like OTP, TNB or polymers) exhibit the
above scaling. According to the available experiments, an
isomorph-invariant expression of the master curve of the scaled data
is not distinguishable from a simpler not-invariant
expression. Instead, the latter agrees better with the simulations of
a wide class of model polymers. The cage rattling of glycerol is
contributed by free-volume over about ten decades in relaxation times.
Both PALS and neutron scattering experiments show that, in order to
evidence the scaling, the observation time of the fast dynamics must
be shorter than the time scales of the relaxation processes.

\section*{Acknowledgement}
Discussions with J.~Barto\v{s} and K.L. Ngai are gratefully acknowledged.

\clearpage
{\bf TABLES}
%
%
\begin{center}
  \begin{scriptsize}
    \ctable[
      caption = {Details about the systems in Fig.\ref{figure1}
        (arranged in order of increasing fragility) and the MD
        simulations used to derive Eq.\ref{eq:scaledparabola}.  The
        experimental structural relaxation time $\tau_\alpha $ is
        drawn by dielectric spectroscopy apart from $\mathrm{B_2O_3}$
        and CKN. $\tau_0$ and $\eta_0$ are the conversion factors
        between the actual time and viscosity units and the
        corresponding MD units, respectively. Note that, apart from
        $\mathrm{B_2O_3}$ and CKN, $\log \tau_0$ and $\log \eta_0$
        cover the narrow ranges $-1> \log \tau_0>-2$ and $-11> \log
        \eta_0 >-12$. MSD is drawn by Incoherent Neutron Scattering
        (INS) or M\"{o}ssbauer Spectroscopy (MS).  The table lists {
          the approximate observation time of the fast caged dynamics
          by the experiment $\Delta t$}, the MSD at the glass
        transition $\langle u_g^2 \rangle$ (in $\mathrm{\AA}^{2}$) or,
        equivalently, the Lamb-M\"{o}ssbauer factor $-\ln f_g$.},
      label = {tab:detailed},
      pos = h,
      center,
    ]{crclrcl}{\tnote[$\dag$]{$\mathrm{Zr_{46.8}Ti_{8.2}Cu_{7.5}Ni_{10}Be_{27.5}}$}
      \tnote[$\ddag$]{Data aggregated from different techniques.}
    }
           { \toprule 
             \multirow {3}{*}{\bf System} &
             \multicolumn{3}{c}{$\tau_\alpha,\ \eta$}&
             \multicolumn{3}{c}{MSD}\\
             \cmidrule(rl){2-4}\cmidrule(rl){5-7}
             &\multirow{2}{*}{quantity}&\multirow{2}{*}{$\log \tau_0$,$\log \eta_{ 0}$}& 
             \multirow{2}{*}{ref.}&\multirow{2}{*}{technique $(\Delta t)$}& 
             \multirow{2}{*}{$\langle u_g^2 \rangle$,$-\ln f_g$}&\multirow{2}{*}{ref.}\\
             & & & & & & \\
             \cmidrule(rl){1-1} \cmidrule(rl){2-2} \cmidrule(rl){3-3} \cmidrule(rl){3-3} \cmidrule(rl){4-4} \cmidrule(rl){5-5} \cmidrule(rl){6-6} \cmidrule(rl){7-7}    
             MD & 
             $\tau_\alpha$ & 0 & \cite{OurNatPhys,OttochianEtAl09}  & 
             MD & 0.01667 & \cite{OurNatPhys,OttochianEtAl09}\\ 
             $\mathrm{SiO_2}$ &
             $\eta$ & -2 & \cite{eta_silica} &
             INS $(40\textrm{ps})$& 0.081 & \cite{tesi_silica}\\
             $\mathrm{GeO_2}$ &
             $\eta$ & -1 & \cite{SippJCNS2001} &
             INS $(0.4\textrm{ns})$ & 0.191 & \cite{CaponiPRB09}\\
             $\mathrm{B_2O_3}$ &
             $\eta$ & +2.2 & \cite{b2o3_3} &
             INS $(40\textrm{ps})$& 0.065 & \cite{b2o3_2}\\
             $\mathrm{B_2O_3}$ &
             $\tau_\alpha$\tmark[\ddag] & -8.4 & \cite{b2o3_1} &
             INS $(40\textrm{ps})$& 0.065 & \cite{b2o3_2}\\
             V4\tmark[$\dag$] alloy &
             $\eta$ & -1 & \cite{Zr_1} &
             MS $(0.1\!\,\mu\textrm{s})$& 0.885 & \cite{Zr_2} \\
             Glycerol & 
             $\eta$ & -1 & \cite{gly_eta} &
             INS $(0.4\textrm{ns})$& 0.022 & \cite{gly2}\\
             1,4 PI & 
             $\tau_\alpha$ & -12 & \cite{PI_1} &
             INS $(4\textrm{ns})$& 0.427 & \cite{PI_2} \\
             TNB &
             $\eta$ & -2 & \cite{TNB} &
             INS $(0.4\textrm{ns})$& 0.315 & \cite{Ngai00} \\
             Fe+DBP &
             $\tau_\alpha$ & -11 & \cite{Fe_2} &
             MS $(0.1\!\,\mu\textrm{s})$& 3.15 & \cite{Fe_2}\\
             Fe+DBP &
             $\eta$ & -2 & \cite{Fe_3} &
             MS $(0.1\!\,\mu\textrm{s})$& 3.05 & \cite{Fe_2}\\
             OTP & 
             $\tau_\alpha$ & -11 & \cite{Ngai00} &
             INS $(0.4\textrm{ns})$& 0.215 & \cite{tolle}\\
             OTP &
             $\eta$ & -1 & \cite{tolle} &
             INS $(0.4\textrm{ns})$& 0.232 & \cite{tolle}\\
             Selenium &
             $\eta$ & -1.66 & \cite{Buchenau92} &
             INS $(20\textrm{ps})$& 0.155 & \cite{Buchenau92}\\
             CKN &
             $\tau_\alpha$\tmark[\ddag] & -13.5 & \cite{SidebottomJCP89} &
             INS $(10\textrm{ps})$& 0.132 & \cite{KartiniPRB96,Ngai97}\\
             1,4 PBD &
             $\tau_\alpha$ & -11 & \cite{PDB_zorn} &
             INS $(4\textrm{ns})$& 0.102 &\cite{frick1,frick2}\\
             a-PP &
             $\tau_\alpha$ & -11.5 & \cite{aPP_1} &
             INS $(0.2\textrm{ns})$& 0.13 & \cite{aPP_2} \\
             PMMA &
             $\tau_\alpha$ & -11.5 & \cite{PMMA_1} &
             INS $(5\textrm{ns})$& 1.1 & \cite{PMMA_2}\\
             PVC &
             $\tau_\alpha$ & -11 & \cite{McCrumEtAl} &
             INS $(5\textrm{ns})$& 0.51 & \cite{PMMA_2}\\
             \bottomrule  \\
           } 
  \end{scriptsize}
\end{center}

\clearpage
{\bf Figure captions}
\begin{figure}[h]
  \begin{center}
  \end{center} 
  \caption{\label{figure1} Reduced relaxation time and viscosity vs
    reduced MSD factor ($ \langle u^2_g \rangle = \langle u^2
    (T_g)\rangle$). The numbers in parenthesis denote the fragility
    $m$. The black curve is Eq.\ref{eq:scaledparabola}. The colored
    curves bound the accuracy of Eq.\ref{eq:scaledparabola}
    \cite{OurNatPhys,OttochianEtAl09}. Refer to Table
    \ref{tab:detailed} for further details about the experiments.}
\end{figure}
\begin{figure}[h]
  \begin{center}
  \end{center} 
  \caption{\label{figure2} Reduced relaxation time (CKN) and viscosity
    (glycerol) vs the reduced MSD. The other curves have the same
    meaning of Fig. \ref{figure1}.  The numbers in parenthesis denote
    the fragility $m$. Other parameters are listed in Table
    \ref{tab:detailed}. }
\end{figure}
\begin{figure}[h]
  \begin{center}
  \end{center} 
  \caption{\label{figure3} Test of Eq. \ref{u2proptotau3} for glycerol
    by using two different data sets for $ \tau_3$ ,
    i.e. ref. \cite{SolesPRL01} (Gly1) and
    ref.\cite{Bartos_Gly_JNCS05} (Gly2).  The plots of the reduced
    glycerol viscosity vs. the reduced time $\tau_3/ \tau_{3 g}$ (
    $\tau_{3 g} = \tau_3 (T_g)$ ) are compared with the master curve
    Eq.\ref{eq:scaledparabola} (black curve). The other curves have
    the same meaning of Fig. \ref{figure1}. $\tau_0$ and $\eta_0$ are
    given in Table \ref{tab:detailed}.  $\log \eta_0' = \log \eta_0 -
    1$. Deviations are observed when the free-volume fluctuations
    become fast with respect to the PALS timescale, i.e. $\tau_\alpha
    \lesssim \tau_3 \sim 2\;\textrm{ns} $. }
\end{figure}
\begin{figure}[h]
  \begin{center}
  \end{center} 
  \caption{\label{figure4} Reduced glycerol viscosity vs. the reduced
    MSD taken at two different timescales $\Delta t = h / \Delta E$
    depending on the energy resolution of the neutron scattering
    experiment $\Delta E$ ( IN13 and IN16 data from refs.\cite{gly2}
    and \cite{BusselezJCP2009}, respectively).  All the curves have
    the same meaning of Fig. \ref{figure1}. $\tau_0$ and $\eta_0$ are
    given in Table \ref{tab:detailed}.  The arrows point to states
    with the indicated $\tau_\alpha$ values.}
\end{figure}
\clearpage
{\bf Figures}
    
\begin{figure}[h]
  \begin{center}
    \includegraphics[width= 0.80\linewidth]{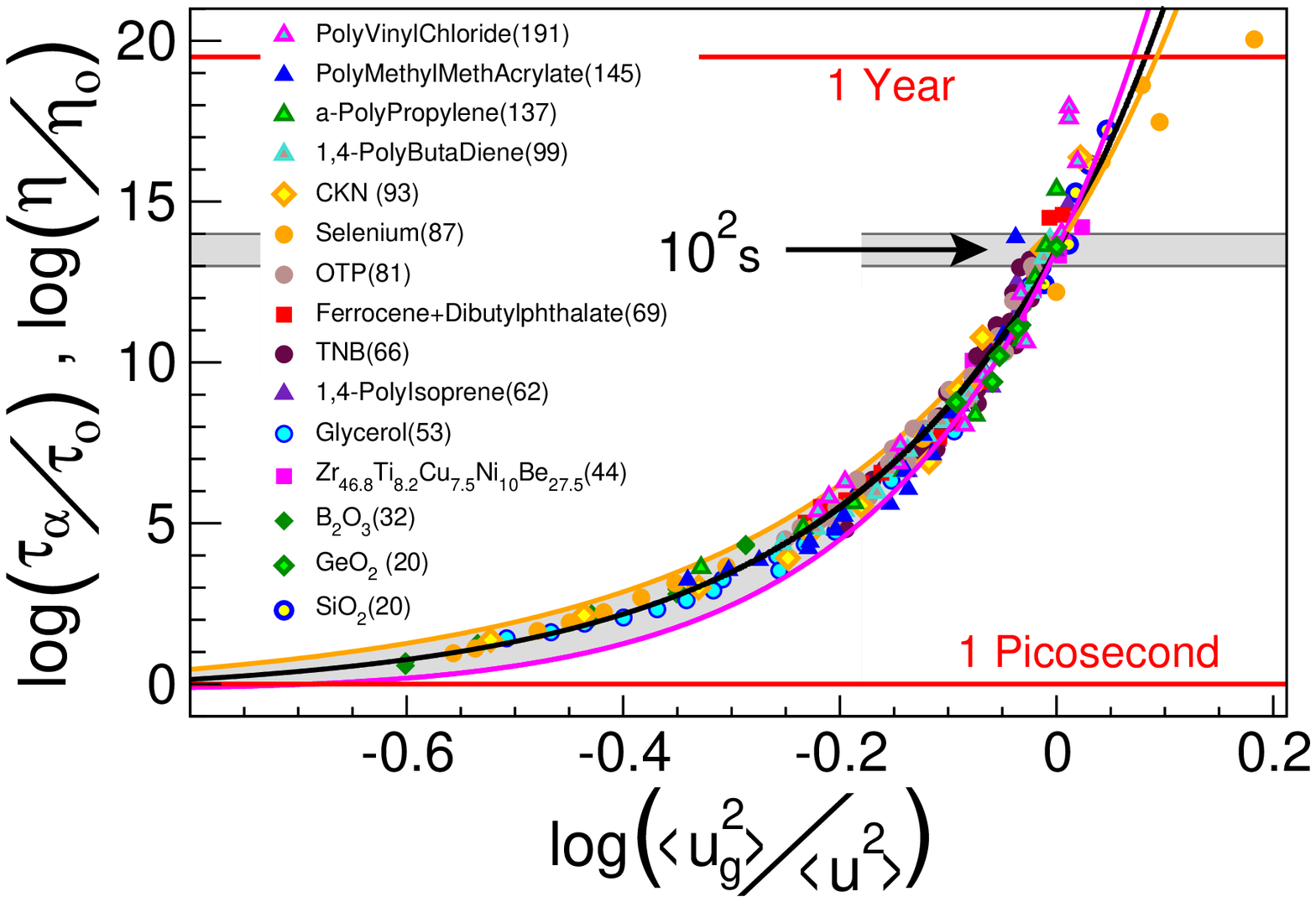}
  \end{center} 
\end{figure}
{\bf FIGURE 1}
\clearpage
\begin{figure}[h]
  \begin{center}
    \includegraphics[width=0.80\linewidth]{figure2.eps}
  \end{center} 
\end{figure}
{\bf FIGURE 2}
\clearpage
\begin{figure}[h]
  \begin{center}
    \includegraphics[width=0.80\linewidth]{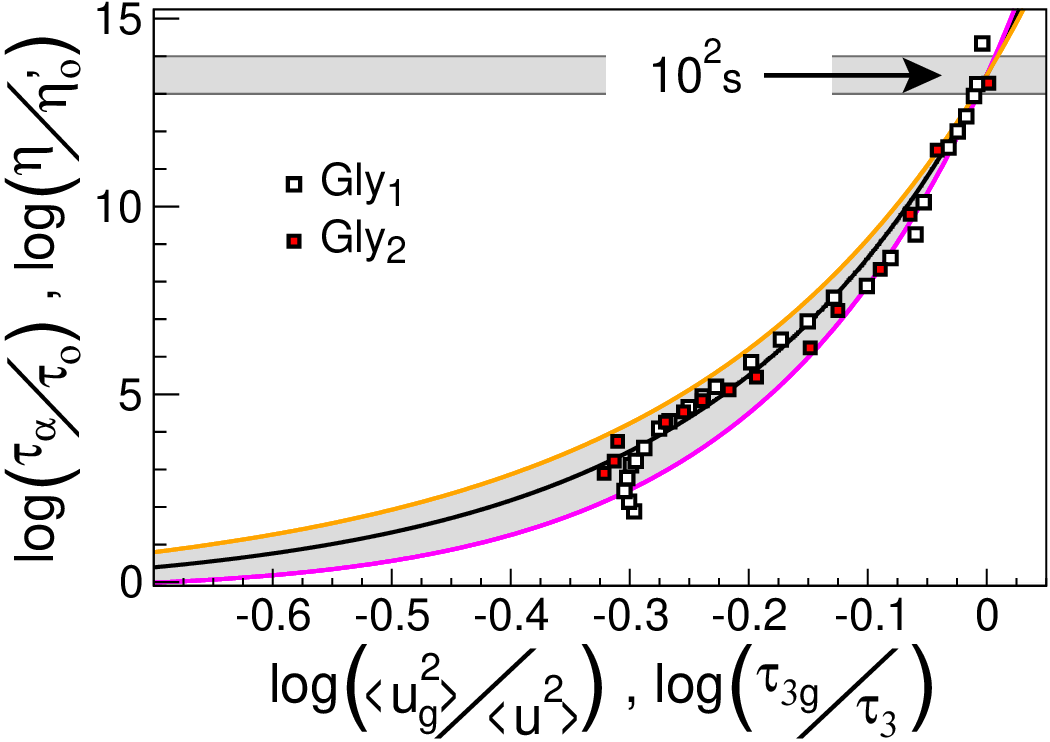}
  \end{center} 
\end{figure}
{\bf FIGURE 3}
\clearpage
\begin{figure}[h]
  \begin{center}
    \includegraphics[width=0.80\linewidth]{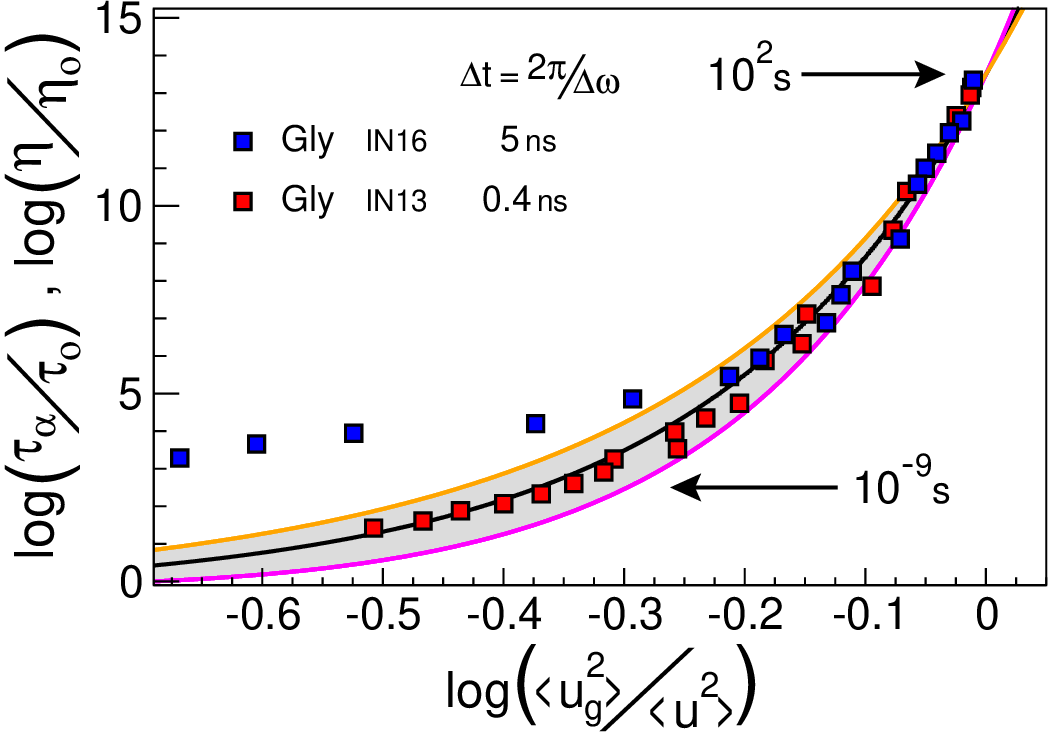}
  \end{center} 
\end{figure}
{\bf FIGURE 4}


\begin{thebibliography}{53}
\bibitem[41]{NoteAndalo}
  In a system with $N$ particles and $\rho$ density, two system states, say $(
  {\bm r}_1, {\bm r}_2, \dots, {\bm r}_N)$ and $( {\bm r}'_1, {\bm r}'_2,
  \dots, {\bm r}'_N)$, are isomorphic under the prescription that, if
  $\tilde{{\bm r}}_i = \tilde{{\bm r}}'_i$, $i=1,\dots, N$ ($\tilde{{\bm r}}_i
  \equiv \rho^{1/3} {\bm r}_i$), then they have proportional configurational
  NVT Boltzmann factors \cite{DyreIsomorphsJCP09}.
\end{thebibliography}
\end{document}